\pdfoutput=1
\documentclass[letterpaper,10pt,twocolumn]{article}

\usepackage{authblk}
\usepackage[english]{babel}
\usepackage[utf8x]{inputenc}
\usepackage{amsmath}
\usepackage{graphicx}
\usepackage[colorinlistoftodos]{todonotes}
\usepackage{mathtools}
\usepackage{hyperref}

\RequirePackage{amsmath,amsfonts,amssymb}
\RequirePackage[T1]{fontenc}
\RequirePackage{changepage}
\RequirePackage[left=48pt,%
                right=42pt,%
                top=46pt,%
                bottom=60pt,%
                headheight=15pt,%
                headsep=10pt,%
                letterpaper,twoside]{geometry}
\newcommand*{\TitleFont}{%
      \usefont{\encodingdefault}{\rmdefault}{b}{n}%
      \fontsize{18}{20}%
      \selectfont}

\title{\TitleFont Reconstruction of a conic-section surface from autocollimator-based deflectometric profilometry}

\author[1,2,*]{Samantha J. Thompson}
\author[3]{Richard Lang}
\author[1]{Paul Rees}
\author[1]{Gareth W. Roberts}

\affil[1]{OpTIC Glynd\^{w}r, Glynd\^{w}r University, Ffordd William Morgan, St. Asaph, UK}
\affil[2]{Astrophysics Group, Cavendish Laboratory, University of Cambridge, JJ Thomson Avenue, Cambridge, UK}
\affil[3]{Private residence, Hertfordshire, UK}

\affil[*]{Corresponding author: sjthompson@cantab.net}

\date{}

\begin{document}

\maketitle

\begin{abstract}
We present a description of our method to process a set of autocollimator-based deflectometer 1-dimensional line-scans taken over a large optical surface and reconstruct them to a best-fit conic-section surface.  The challenge with our task is that each line-scan is in a different (unknown) coordinate reference frame with respect to the other line-scans in the set.  This problem arises due to the limited angular measurement range of the autocollimator used in the deflectometer and the need to measure over a greater range; this results in the optic under measurement being rotated (in pitch and roll) between each scan to bring the autocollimator back into measurement range and therefore each scan is taken in a different coordinate frame.  We describe an approach using a $6N+2$ dimension optimisation (where $N$ is the number of scan lines taken across the mirror) that uses a gradient-based non-linear least squares fitting combined with a multi-start global search strategy to find the best-fit surface.  Careful formulation of the problem is required to reduce numerical noise and allow the routine to converge on a solution of the required accuracy.
\end{abstract}

\bigskip
{\bf Keywords (ociscodes):} 120.3940 Metrology; 120.6650 Surface measurements, figure; 220.1250 Aspherics.

\bigskip
\textcopyright 2016 Optical Society of America. One print or electronic copy may be made for personal use only. Systematic reproduction and distribution, duplication of any material in this paper for a fee or for commercial purposes, or modifications of the content of this paper are prohibited.  The version of this paper as published in Applied Optics can be viewed here:\url{http://dx.doi.org/10.1364/AO.55.002827}

\section{Introduction}
\begin{figure}[htbp]
\centering
\includegraphics[width=\linewidth]{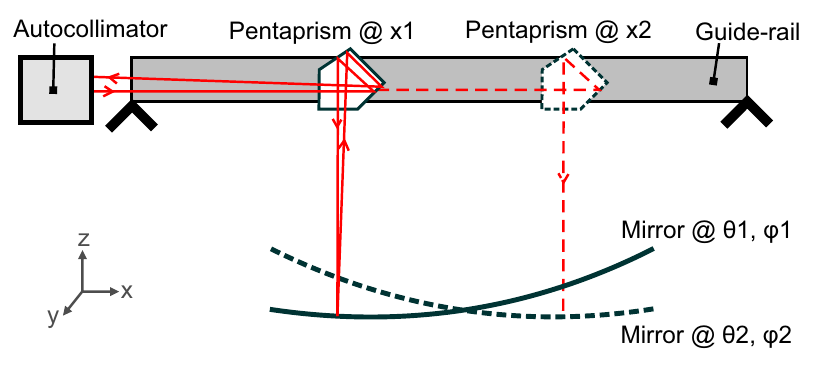}
\caption{Diagram of an autocollimator-based deflectometer set-up.  The pentaprism traverses along x relaying the optical beam from the autocollimator to the mirror surface and back again.  Motion in y is provided via the polishing machine that the entire set-up is mounted on.  The large mirror depicted has a slope range that is larger than the angular measurement range of the autocollimator so the mirror must be repositioned in pitch and roll ($\theta 1$ to $\theta 2$ and $\phi 1$ to $\phi 2$) to enable measurements over the entire surface (e.g. dotted path).}
\label{fig:nom}
\end{figure}

The OpTIC NOM \cite{Atkins11,Pearson15} is a deployable, long-arm, non-contact profilometer consisting of a scanning pentaprism and a digital autocollimator; this type of profilometer is also known as a deflectometer.  A deflectometer can be used as a non-contact probe to provide accurate measurement of the height profile across an optical surface.  The pentaprism traverses along a stiff linear guide-bar above an optical surface, relaying the autocollimator beam to provide a set of surface slope (angle) measurements along that path.  Pentaprism position information is provided via a linear encoder and so a height profile of the optical surface can be calculated.  A diagram of the set-up is given in Fig. \ref{fig:nom}.  The instrument is being used to confirm the base radius of curvature ($R$) and conic constant ($k$) of prototype European Extremely Large Telescope (E-ELT) segments that are being manufactured at OpTIC Glynd\^{w}r.

This profilometry technique originated in the synchrotron community \cite{qian95,alcock10} to provide accurate measurements of X-ray focussing optics.  These mirrors for synchrotron X-rays generally fall under two categories: (1) long (up to 1.5 m) and narrow (a few 10's of mm) with very long focal lengths (i.e. very large radii of curvature with $R$ from 100's of metres up to a few kilometres) and (2) nano or micro-focussing X-ray mirrors that can have higher surface curvature (i.e. $R \sim 10$m) but are only $\sim 100$mm long.  In both these cases the change in slope over the surface of the mirrors is usually < 10 milliradians so the limited angular measurement range of the digital autocollimator is not an issue and angular "stitching" techniques are not required in the data analysis.  

For our purpose we do not require the extreme accuracy and precision that is needed for X-ray optics, but we need to measure large optical surfaces (e.g. > 1 m diameter areas) with slope variations that exceed the range of most commercially available, sub-arcsecond accuracy, digital autocollimators.  Also, in contrast to the X-ray optic community, the OpTIC NOM measurements are subjected to a significantly higher level of environmental noise due to the fact that the instrument is portable and used in-situ over a robotic polishing machine.

This paper describes a numerical method that can be used to process sets of deflectometer line data that have been taken over an optical surface, where the individual line scans in that data set have been taken in difference coordinate frames, and fit an aspheric (conic-section) surface to them.  Simulated datasets have been used to test the fitting method in order to determine the accuracy of the method and any limiting behaviour. The software has been implemented using M\textsc{atlab} \cite{matlab}.

\section{Description of the task}
\label{sec:description}
For a primary mirror (M1) as large as that proposed for the E-ELT, it is not feasible to manufacture the mirror in one continuous piece; it must be constructed from many smaller mirrors that are of a size more practical to manufacture, measure and transport.  The primary aperture of the E-ELT at the time of this study was a 42 m diameter ellipsoid.  The surface form of the primary mirror is given by the formula: 
\begin{equation}
z=  \frac{(1/R)(x^2+y^2)}{1+ \sqrt{1-(1+k)(1/R)^2 (x^2+y^2)}}
\label{eq:ellipform}
\end{equation}
(Ref: \cite{eso3}, p. 16, the Ellipsoid formula).  Where $R$ is the base radius of curvature, $k$ is the conic constant and $x$, $y$, $z$ are the coordinates in the M1 (primary mirror) reference frame.

The following production and measurement accuracies were specified by ESO \cite{eso3}:
\begin{itemize}	
\item The first polished segment should have a base radius of curvature and conic constant which satisfy $R1 = 84000$ mm $\pm 200$ mm and $k1 = -0.993295-(R1 -84000) \times 3.206302 \times 10^{-7}$.
\item The accuracy of the knowledge of the nominal radius of curvature, $R1$ shall be better or equal to 14 mm RMS.
\item The maximum allowable surface error (before removal of low and mid-spatial frequency terms) over the useful area of any prototype segment is 50 nm RMS (40 nm goal).
\end{itemize}

The 42m primary consists of 1148 hexagonal mirror segments of approximately 1.4 m diameter.  The set of prototype segments to be manufactured and measured were a group of 7 segments towards the outer edge of the primary mirror assembly, i.e. each mirror segment is an off-axis ellipsoid.  

One of the challenges in manufacturing the primary mirror is that each individual segment must appear to have been “cut-out” from this larger ellipsoid - each segment must be made to the same $R$ and $k$ (within tolerance) and this surface form must be correctly aligned to the geometry of the segment (which are slightly irregular hexagons).  Interferometric testing provides accurate surface-form maps in relation to the geometry of the segment, however it cannot provide a direct measurement of $R$.  A more direct measure (i.e. one not requiring a reference optic) can be provided using a deflectometer.  Therefore, the primary task of the OpTIC NOM is to provide a confirmation of the base radius of curvature ($R$) and conic constant ($k$) of prototype E-ELT mirror segments during the end stages of optical polishing.  

\begin{figure*}[hbt]
\centering
\includegraphics[width=\linewidth]{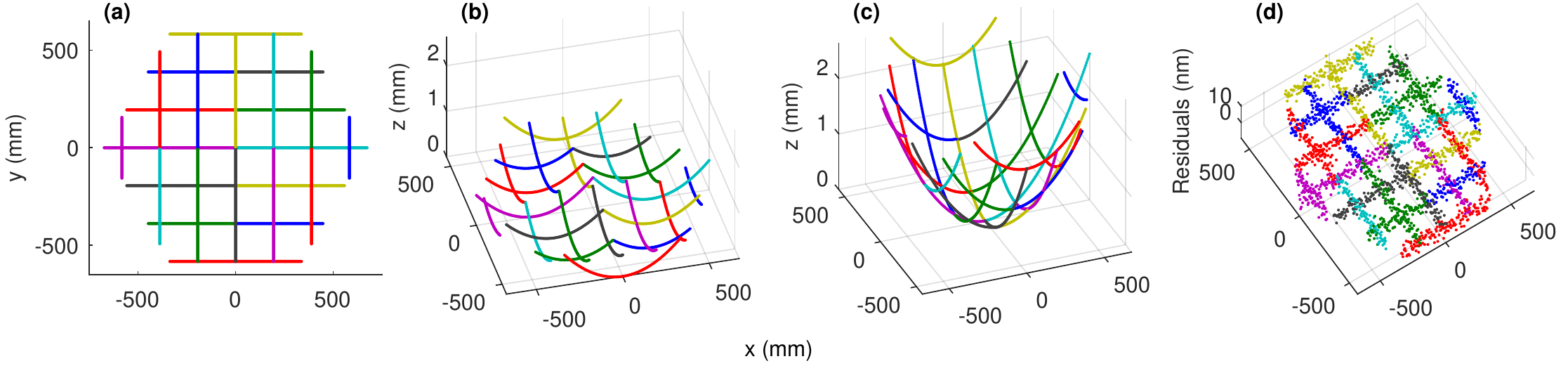}
\caption{(a) Plot showing a set of scan lines suitable for sampling the surface of a 1.4m diameter hexagonal mirror.  The colours are used only to highlight the individual scan lines (24 in total in this set), there is no overlap along the different scan-lines. (b) Due to the data acquisition method described, this is how the scans look when plotted after they have been integrated from angular data to heights ($z$). (c) The same data as in (b) after it has been analysed used the described fitting-optimisation process - the individual line-scans are now co-located on the best-fit conic surface. (d) A plot of the residuals after the best-fit conic surface is subtracted from (c), the RMS of these residuals is consistent with the random noise level in the data.}
\label{fig:scanlines}
\end{figure*}

\subsection{Frames of reference - definitions}

In the following discussion, reference is made to the M1 (primary mirror) coordinate reference frame and the segment (denoted by the segment number: e.g. S1, S4 etc.) coordinate reference frame.  The position and surface form of the segments is defined in M1 coordinates as described by Equation \ref{eq:ellipform}.  When measuring a segment by interferometry, the segment is supported in its gravitationally symmetric position: the mirror segment is oriented so that the normal to the surface at its centre coordinate is pointing vertically upwards (along the z-axis).  In comparison, in the M1 coordinate reference frame, the normal vector of a general segment is not pointing along z; only a segment centred at the centre of M1 would possess this property.

When measuring with the deflectometer the mirror segment is, on average, oriented with its centre-normal pointed vertically upwards.  However, as previously described, the range of surface slope on a segment exceeds the measurement range of the autocollimator and so the segment must be pitched and rolled (see Fig. \ref{fig:nom}) to allow scans to be taken over the entire surface.  The range of these angular re-positionings is of the order of $\pm 0.25$ degrees.  With the segment tilted in various orientations for different line-scans the measurement coordinate frame is no longer co-aligned with the segment coordinate frame.  This makes co-locating all the separate line scans to a single surface - which needs to be a best fit to this data - non-trivial.

Figure \ref{fig:scanlines}(a) gives an example of a set of scan-lines taken over a segment surface, Fig.\ref{fig:scanlines}(b) shows the effect that the tilting method used to overcome the limited measurement range has on the data and Fig.\ref{fig:scanlines}(c) shows how the line scans appear when each scan is properly corrected for z-height offsets and rotations after the optimisation/fitting procedure.

\section{Fitting a set of line-scans to a conic surface}

\begin{figure}[tb]
\centering
\includegraphics[width=\linewidth]{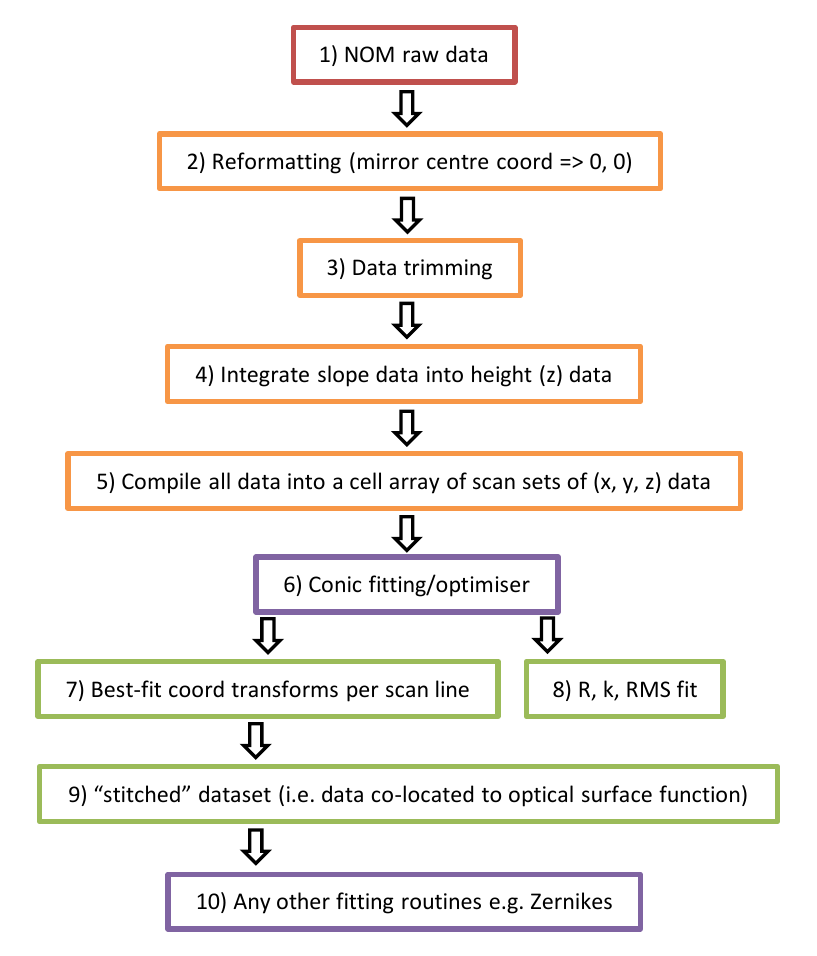}
\caption{A flowchart summarising the key data processing steps for NOM data.  The simulations discussed here start at step (5) in this flowchart using simulated datasets.  For this work, the surface function in (9) is given by the form of the Ellipsoid formula described by Eq.\ref{eq:zref}.  The final step (10) is beyond the scope of this work.}
\label{fig:dataflow}
\end{figure}

A suite of M\textsc{ATLAB} programs have been written to read-in, process and find the best-fit conic surface for the NOM line-scan datasets.  The data flowchart for these processes is shown in Fig. \ref{fig:dataflow}.  As previously described, the challenge with fitting this data to a surface is that due to the limited angular range of the autocollimator, each scan-line taken on the mirror surface is in an unknown different coordinate reference frame and so the fitting routine must translate and rotate the scans to correct for this.  Also, given that the $z$ (height) data of the surface is obtained via integration of slope data (of the form d$z$/d$x$ = tan$\theta$), the absolute $z$ position of each scan is unknown (the unknown integration constant), so the relative $z$ height of all the scans with reference to each other must also be part of the optimisation routine.

For a given set of scan-lines the parameters that require fitting are: $R$ and $k$ (globally) and also 6 parameters per scan (3 translations and 3 rotations).  This means there are $6N + 2$ parameters to fit per set of line-scans (where $N$ is the number of line scans in a set).  For the line-scan pattern shown in Fig. \ref{fig:scanlines}, this means there are $(6 \times 24) +2 = 146$ parameters to optimise in the conic fitting.  To perform this optimisation a trust-region-reflective algorithm is used to determine the minimum of a cost-function.  As expected, a gradient minimisation over so many dimensions is computationally costly and prone to stopping in local minima.  To help the optimiser, “good” initial values of $R$ and $k$ must be provided and boundary limits applied to the individual scan-line transformations.  In addition a global search strategy is used to circumvent the local-minima problem – starting the optimiser with many different sets of initial conditions and selecting the most optimal.

This fitting routine can be run on a standard desktop PC (for this benchmark the processor is an Intel\textsuperscript{\textregistered} Xeon\textsuperscript{\textregistered} CPU E3-1240 v3 3.40 GHz).  As an example of the run-time, a single run of the optimisation/fitting routine (i.e. using only a single $(R, k)$ pair as an initial starting value) on a typical dataset simulated here (24 scan-lines containing 2562 data points in total) which requires 146 parameters to optimise takes $\sim 508$ seconds to converge.

\subsection{The Cost Function}
An important step in setting up this optimisation routine is the formulation of an appropriate cost function, $F$. The quantity that we minimise is the sum of the squared differences between the height of each point in each line scan ($z_i$) and the calculated $z$ of an ellipsoid surface at the same ($x_i$, $y_i$) coordinate positions for a given $R$ and $k$ (called $z_{ref}$ here).

\begin{figure}[htbp]
\centering
\includegraphics[width=\linewidth]{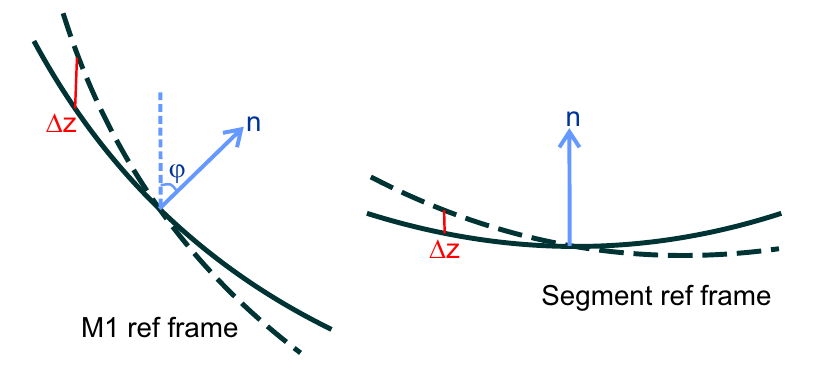}
\caption{Diagram to illustrate how the cost function can be affected depending on the reference frame (reference surface in solid line) in which the cost function is evaluated; angles have been exaggerated to illustrate the issue.  As an example, the centre coordinate in the M1 reference frame for segment S1 = (0, 18470, 2031) mm and in the S1 reference frame the centre is transformed to (0, 0, 0) mm with the normal to the surface at the centre pointing vertically upwards (along the z-axis).}
\label{fig:refframes}
\end{figure}

To enable the comparison between $z_i$ and $z_{ref}$, either the line scans must be transformed into the M1 coordinate frame or the Ellipsoid formula (Equation \ref{eq:ellipform}) transformed into the relevant segment coordinate system. The more reliable of these for the cost function is the latter - a comparison in the segment coordinate frame. One of the reasons for this is that in the segment reference frame the cost function ($z_i$ – $z_{ref}$) is less weighted by the outer scan lines (furthest from the segment centre) for the same angular displacement, another is that the cost function is smoother since most of the different parameter values are closer to zero than in the M1 frame. Figure \ref{fig:refframes} presents an illustration of this issue. 

The transformation of the ellipsoid formula from M1 to segment reference frame uses the pre-defined centre coordinate of the segment under manufacture and the measured centre coordinate determined experimentally from the segment geometry.  The $z$ value of the segment centre, in segment coordinates is set to be $z = 0$.

The form of the conic equation used to calculate $z_{ref}$ is:
\begin{equation}
z_{ref}= \frac{\frac{x^2 + y^2(1 + k \sin^2 \phi)}{R} - 2y \sin \phi} {\splitfrac{\cos \phi (1 - \frac{ky \sin^2 \phi}{R})} {+ (\cos^2 \phi + \frac{2y \sin \phi}{R} - \frac{y^2(1+k) + x^2(1+k \cos^2 \phi)}{R^2})^{0.5}}}
\label{eq:zref}
\end{equation}
where $R$ and $k$ are the ellipsoid parameters, $x$ and $y$ are the coordinates in the segment frame and $\phi$ is the rotation angle around the x-axis that brings the normal vector of the segment at its centre in the M1 frame to a vertical orientation (i.e. along the positive z-axis); this is also illustrated in Fig. \ref{fig:refframes}.  The equation can be formulated in this way because the Ellipsoid formula (Equation \ref{eq:ellipform}) is symmetric about the z-axis, and for the purposes of comparing the theoretical surface with the real data the data can be transformed so that the segment centre lies on the y-axis.

The optimization routine being used implicitly computes the sum of squares as part of the algorithm, so our cost function, $F$, is multi-valued and creates a length $n$ vector of z differences, one for each measured ($X$,$Y$,$Z$) value:
\begin{equation}
\begin{split}
F(R, k, \{ {\theta}x_s, {\theta}y_s, {\theta}z_s, {\delta}x_s, {\delta}y_s, {\delta}z_s : s = 1 \ldots N  \}) \\
= [f_1,\dotsc,f_i,\dotsc,f_n]
\label{eq:F}
\end{split}
\end{equation}
where $N$ is the number of line scans and
\begin{equation}
f_i = z_i - z_{ref}(R,k,x_i,y_i,\phi)
\label{eq:f_i}
\end{equation}
and where $x_i$, $y_i$, $z_i$ are the rotated and translated line scan measurements given by
\begin{equation}
\mathbf{p_i} = \mathbf{T_s}\mathbf{P_i} + \boldsymbol{\delta} \mathbf{_s}     .
\end{equation}
Here
\begin{equation}
\mathbf{p_i} = \begin{pmatrix}x_i \\ y_i \\ z_i\end{pmatrix}, \qquad\mathbf{P_i} = \begin{pmatrix}X_i \\ Y_i \\ Z_i\end{pmatrix}, \qquad
\boldsymbol{\delta}\mathbf{_s} = \begin{pmatrix}{\delta}x_s \\ {\delta}y_s \\ {\delta}z_s\end{pmatrix}
\end{equation}
and
\begin{equation}
\mathbf{T_s} = \mathbf{Tx_s} \mathbf{Ty_s} \mathbf{Tz_s}
\end{equation}
where $X_i$, $Y_i$, and $Z_i$ are the measured coordinate position and height data for point $i$ and $s$ is the line scan number which contains the point $i$. $\mathbf{Tx_s}$,  $\mathbf{Ty_s}$ and $\mathbf{Tz_s}$ are the standard rotation matrices around the $x$, $y$ and $z$ axes respectively for the angle parameters $\theta x_s$, $\theta y_s$, $\theta z_s$ for line scan number $s$.

\subsection{The Optimisation algorithm}
A trust-region-reflective algorithm is used in the optimiser and it is set up as a bounded problem - initial values are allocated to the parameters that require fitting and limits are also defined for those parameters.  The optimisation is based around a gradient minimisation; the cost function gradient is calculated using the central finite differencing method.  For this problem the upper and lower boundary limits were chosen to be symmetric around the initial value, $v_{initial}$, of each parameter $param$:
\begin{equation}
(v_{initial} - \Delta) \leq param \leq (v_{initial} + \Delta)   .
\end{equation}
For the 6 degrees of freedom in positioning each scan line into a common (the segment) coordinate frame the ${\delta}x$, ${\delta}y$, ${\delta}z$ limits are set at $\pm 5$ mm and the 3 rotations are limited to $\pm 0.05$ radians; the initial values are set to zero.  These limits have been chosen to be suitable for the measurement set-up as previously described and the expected displacements and rotations of the line scans are well within these boundary limits.  For $R$ and $k$, the global parameters across all scan lines, the initial values, $R_i$ and $k_i$, are set to a good estimate based on other measurements (e.g. interferometric measurements) and the boundary limits in the program are set to $R_i \pm 2000$ mm and $k_i \pm 0.5$.  Section \ref{sec:performance} demonstrates the method used to reach the correct solution even if good initial estimates for $R$ and $k$ are not available.

\begin{figure}[tb]
\centering
\includegraphics[width=\linewidth]{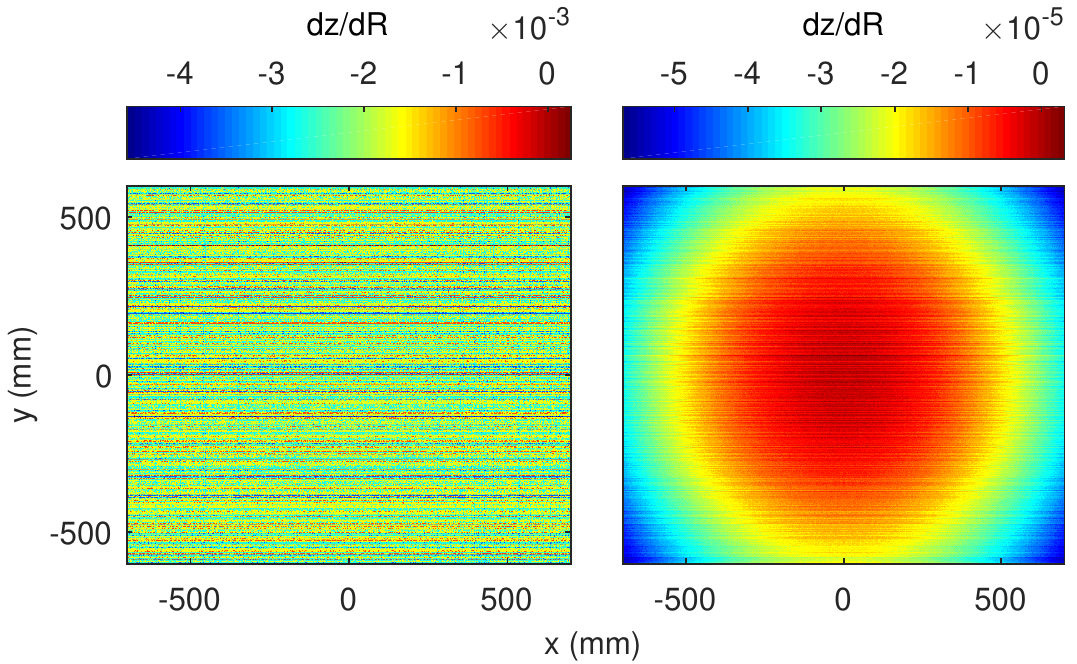}
\caption{Examples of numerical noise in the Jacobian for $R$ (i.e. d$z$/d$R$) over $x$ and $y$.  Both plots show the same Jacobian but calculated using different forms of the quadratic equation solution - the left plot is calculated before the function calculations were optimised for numerical precision and the right plot is after the reformulations to improve numerical noise.  These have been calculated for S1 (near the outer edge of M1), which is the worst case (in terms of the numerical noise) due to the magnitude of the y-coordinates.  Towards the centre of M1, the centre coordinates describing the segment positions are more moderate in magnitude resulting in a smoother Jacobian.}
\label{fig:numnoise}
\end{figure}

\subsection{Numerical issues}
In order to allow the optimisation algorithm to operate efficiently, it is important to present it with a cost function that is as smooth as possible.  Any noise in the cost function causes fluctuating gradient measurements which at best will cause the optimisation to take a longer path to the solution, and at worst may cause it to terminate prematurely in a false local minimum.  An example of problems with numerical noise is shown in Fig. \ref{fig:numnoise}(left plot): the initial formulation of the problem results in such significant loss of numerical precision that there is no discernible gradient for the optimiser to follow and it terminates in an apparent local minimum very rapidly (it hardly moves from the initial starting values).  Figure \ref{fig:numnoise}(right plot) shows the same gradient function after applying several improvements as described below.  Numerical noise in the cost function is still present, but it is now significantly below the level of the actual function evaluation values, and low enough that the derived gradients can be followed by the optimiser.  

\subsubsection{Cost function formulation}
A standard formulation of equations in the cost function  can result in a significant loss of numerical precision due to cancellation between terms of the ellipsoid equation.  By reformulating the calculation of the surface using a different form of the quadratic equation (\cite{recipes}, Ch.5, pp.227) the numerical accuracy of the calculation is vastly improved. The form of the cost function ($z_i - z_{ref}$)  used is calculated from Equation \ref{eq:zref}. In addition, the order of arithmetic operations throughout the cost function was carefully analysed in the context of the expected magnitudes of the inputs and improved where possible to maximise the precision of the cost function.

\subsubsection{Numerical gradient method}
The choice of how the gradient is calculated has a significant effect on the solution convergence - the central (or symmetric) finite differencing method has been chosen which is computationally more costly than the forward finite differencing method but the central method has higher numerical precision (\cite{recipes}, Ch.5.7, pp.229). Use of the forward method results in a solution of much lower accuracy and is insufficient for the solution accuracy required here.

\subsubsection{Optimisation step size}
The default step size for Jacobian calculations in Matlab is quite small (1e-8), although it can be set to any value from 0 to inf.  Increasing the step size will sample over a size greater than the granularity of the numerical noise features and so a smoother function is produced which is then less likely to present the optimiser with false (noise based) local minima.  However, the step size also needs to be minimised to provide an accurate estimation of the gradient function and ensure that the optimiser will converge on a solution of the required accuracy.

With the initial cost function formulation, a step size of 1e-3 ($\sim$1e9*eps($R_0$)), where $R_0 = 84000$, was needed to produce the required smoothness, but this results in very poor gradient estimates.  Machine epsilon (eps) is the approximate relative error due to rounding in floating point arithmetic of a 64 bit double representing that value.  With the improved cost function, a step size of 1e-7 ($\sim$1e4*eps($R_0$)) could be chosen which provides a good balance between noise reduction and accuracy of estimation.

\subsubsection{Parameter normalisation}
The optimisation algorithm allows only a single value for the gradient estimation step size, $h$, to be chosen to apply to all the parameters.  The parameters in our problem differ by several orders of magnitude, so for the optimisation process the inputs are normalised such that their bounds map to [-1 1].  A step size of 1e-7 is then suitable for the entire n-dimensional gradient space minimisation.

\subsubsection{Global search}
To guard against solutions that may have converged in a local minimum rather than the global minimum a global search strategy is used, starting the optimisation routine using many sets of different initial conditions and comparing the results.  An efficient way to implement this strategy is to generate a randomised Sobol sequence \cite{sobol76} over the required n-dimensional search space.  By using $2^m$ (where $m$ is the number of sets of initial parameters) points to generate the Sobol set the points will be uniformly distributed over the required search space.  For this work we have generally been using this to generate a set of $(R, k)$ initial values to verify the solution convergence.

\begin{figure}[bt]
\centering
\includegraphics[width=\linewidth]{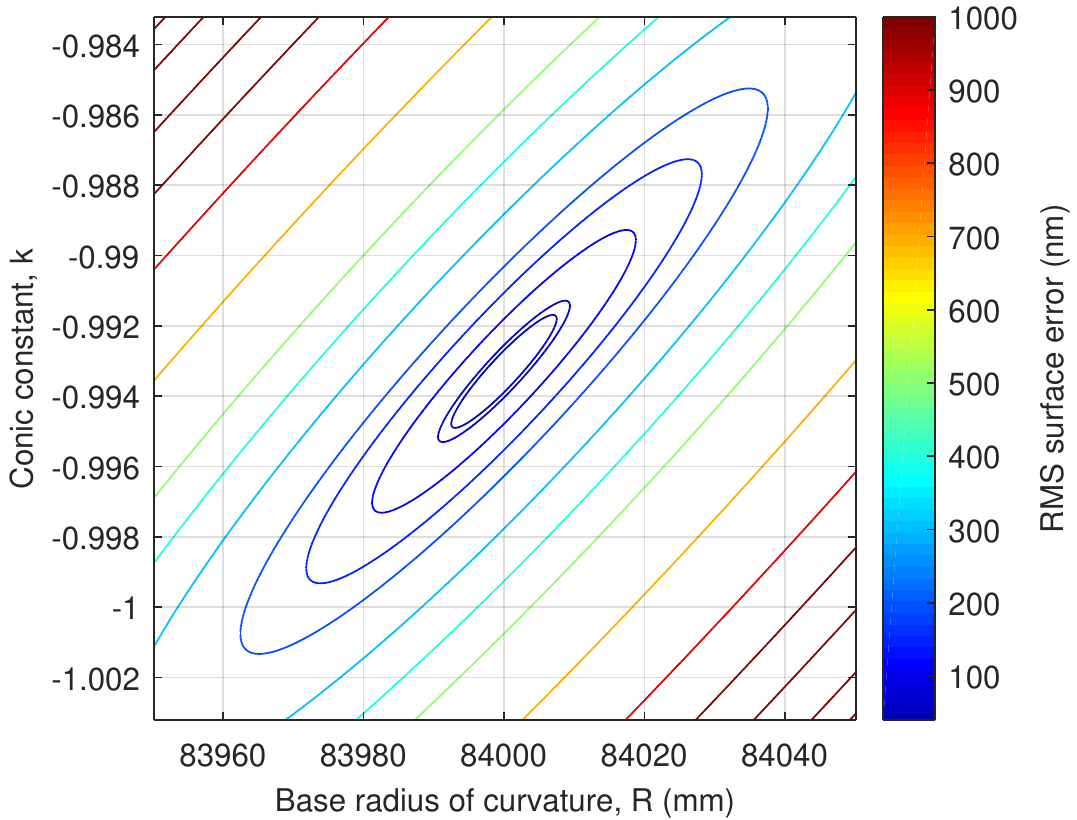}
\caption{Contour plot of the $R$, $k$ parameter space that occupies the same RMS surface error – contours map solutions having the same RMS surface error.  The inner 10 contour levels are (in nm): 40, 50, 100, 150, 200, 300, 400, 500, 700 and 900.  Given $R_0 = 84000$mm and $k_0 = -0.993295$ for a “perfect” first segment, the ESO allowable surface error of 50nm RMS for subsequent segments corresponds to the plotted elliptical boundary limits in ($R$ (mm), $k$) of (83990.7, -0.995295) and (84009.3, -0.991291).}
\label{fig:contours}
\end{figure}

\section{Testing methodology}

For the real data, sources of error such as environment noise (vibration, dust, air turbulence), alignment errors, encoder errors etc. lead to measurement errors in the autocollimator slope angle.  In principle these slope errors could be estimated using classical error propagation, but this alone does not give us the error on the parameters we are trying to determine: $R$ (base radius of curvature) and $k$ (conic constant).  The slope angles are converted into gradients and integrated to $z$ values.  The cumulative integration error could be estimated and would vary depending on the numerical integration method. 

As previously described, the Ellipsoid formula (Equation \ref{eq:ellipform}) is fitted to the set of line scans.  Due to the limited range of the autocollimator, line scans covering the full extent of the segment can only be obtained by tip, tilt and rotation of the segment on its support structure.  Due to this, the parameters in the fitting routine include $R$ and $k$, but also separate spatial translation and rotation parameters for each line scan, to allow for the fact that we only approximately know their relative locations.  An analytical propagation of errors to the parameters $R$ and $k$ is not possible in this situation (a non-linear iterative fitting process); the error in $R$ and $k$ from this fitting process is better estimated using Monte Carlo methods and will depend on the optimisation algorithm and the exact cost function formulation as well as the errors in $x$ and $z$.  In theory, errors in our estimates for the translations/rotations of the line scans can be corrected by the optimiser but in practice they will also affect the ability of the algorithm to fully converge.

To test the behaviour and accuracy of the fitting routine simulated NOM scans are generated.  As shown in Fig. \ref{fig:scanlines}(a), a set of scan coordinates defined in $x$ and $y$ in the segment reference frame are passed to the NOM scan simulator along with the surface specification in $R$ and $k$, the centre coordinate in the M1 reference frame and the 1-sigma level of z-noise to add to the generated $z$ values.  Based on a measurement procedure of ensuring each scan is centred within the surface slope range of the autocollimator, the data for each scan line is then rotated in the $x$ and $y$ axes around the centre point of the scan line so that the $x$, $y$ slopes at this centre point are zero.  Note that the specification of the $x$, $y$ coordinates of the scan lines are chosen to ensure that the slope measurement along its length does not exceed the total measurement range of the autocollimator.  Realistic levels of random noise on $z$ are chosen based on real data acquired with the OpTIC NOM.  The simulated dataset then looks like Fig. \ref{fig:scanlines}(b) and these data are passed to the fitting routine.  In terms of the whole process shown in Fig. \ref{fig:dataflow}, the simulated dataset comes in at step 5 and represents the data as if it has gone through the previous steps and the data taking process (not indicated).

For all the performance tests presented here the simulated datasets were generated in the format as shown in Fig. \ref{fig:scanlines}(a) (i.e. a scan pattern consisting of a grid of 7 lines across each of the x and y axis to produce 24 scan lines that are within the measurement range of the NOM autocollimator).  The data points within a scan line are separated by 5 mm.

Since the data used in these tests are simulated, the $R$ and $k$ is known and the level of random noise injected into the data is known.  Sanity checks on the reconstructed/fitted data are made by checking that the final value of the cost function (which is basically a sum of the residuals) corresponds to the level of noise that was injected (i.e. an RMS of the residuals should be close to the 1-sigma level of noise that was used).  If the residuals are higher than expected it is usually because the optimiser has terminated too early. This can happen if the initial parameters were beyond the bounds of the optimiser (see Section \ref{sec:outbounds}), or the initial values were located unfavourably in the gradient-space so that the route taken by the optimiser needed more steps than the allocated maximum number of iterations; it is for these reasons that a multi-start strategy is best used for unknown data.  The reconstructed data (as per Fig. \ref{fig:scanlines}(c)) as well as the residual plot (Fig. \ref{fig:scanlines}(d)) can also be plotted to double-check that the scan lines have been sensibly located on the best-fit surface and that there is nothing unexpected in the residuals (i.e. that they are consistent with the expected noise).

Finally, in order to determine whether a solution has converged within the limits of the allowed RMS surface error a contour plot of RMS surface error was derived over the $R$, $k$ parameter space.  The ESO allowable surface error is defined to be 50nm RMS \cite{eso3}.  For segment S1, whose centre coordinate is positioned near the outer edge of the primary mirror this contour plot is given in Fig. \ref{fig:contours}.  The RMS surface error is calculated based on the 2562 data points in the line scan pattern (Fig. \ref{fig:scanlines}(a)) of the typical simulated scan set used, with the minimum surface error (injected $z$ noise only) set at $R_0 = 84000$ mm and $k_0 = -0.993295$.  The RMS surface error versus $(R, k)$ varies with position on M1, so a different contour map must be calculated for each segment set to enable an accurate evaluation of the $R$, $k$ solution convergence.  An example of this for a segment closer to the inner edge (named S12 here) is given in Section \ref{sec:innerseg}.  Comparing the $R$, $k$ limits on the 50 nm RMS contour for S12 gives: Rlimits = 83991.1 – 84008.9 mm (diff = 17.8 mm) and klimits = -1.009890 - -0.976677 (diff = 0.033213).  The kdiff for S1 is 0.00392392 (an order of magnitude smaller) and the Rdiff is 18.6 mm.

\section{Performance}
\label{sec:performance}

When using this analysis method on a real data-set there are certain unknowns that the analysis must take into account and be robust to.  It is assumed that approximate values are known for the key parameters that define the shape of the surface (e.g. base radius of curvature and conic constant), so that these can be used as reasonable initial values in the optimisation process.  However, even if there is no prior knowledge of these parameters a wider search strategy can be employed using more steps in the process; the time taken to reach a satisfactory solution is increased accordingly.

A number of different data sets were created to test the fitting method and these are described in the following sub-sections.

The boundary box in the optimiser is set at $R_i \pm 2000$ mm and $k_i \pm 0.5$ (where $R_i$ and $k_i$ are the initial values provided to start the optimiser); so if the initial values fall beyond the real solution $\pm$ the $R$ and $k$ boundary limits then the routine will fail to find the correct solution because it is beyond its search-space.  It is unlikely that the user would ever have this level of uncertainty when providing a good initial value, however an example is given (in Section \ref{sec:outbounds}) of what happens in this case.

\begin{figure}[t!]
\centering
\includegraphics[width=\linewidth]{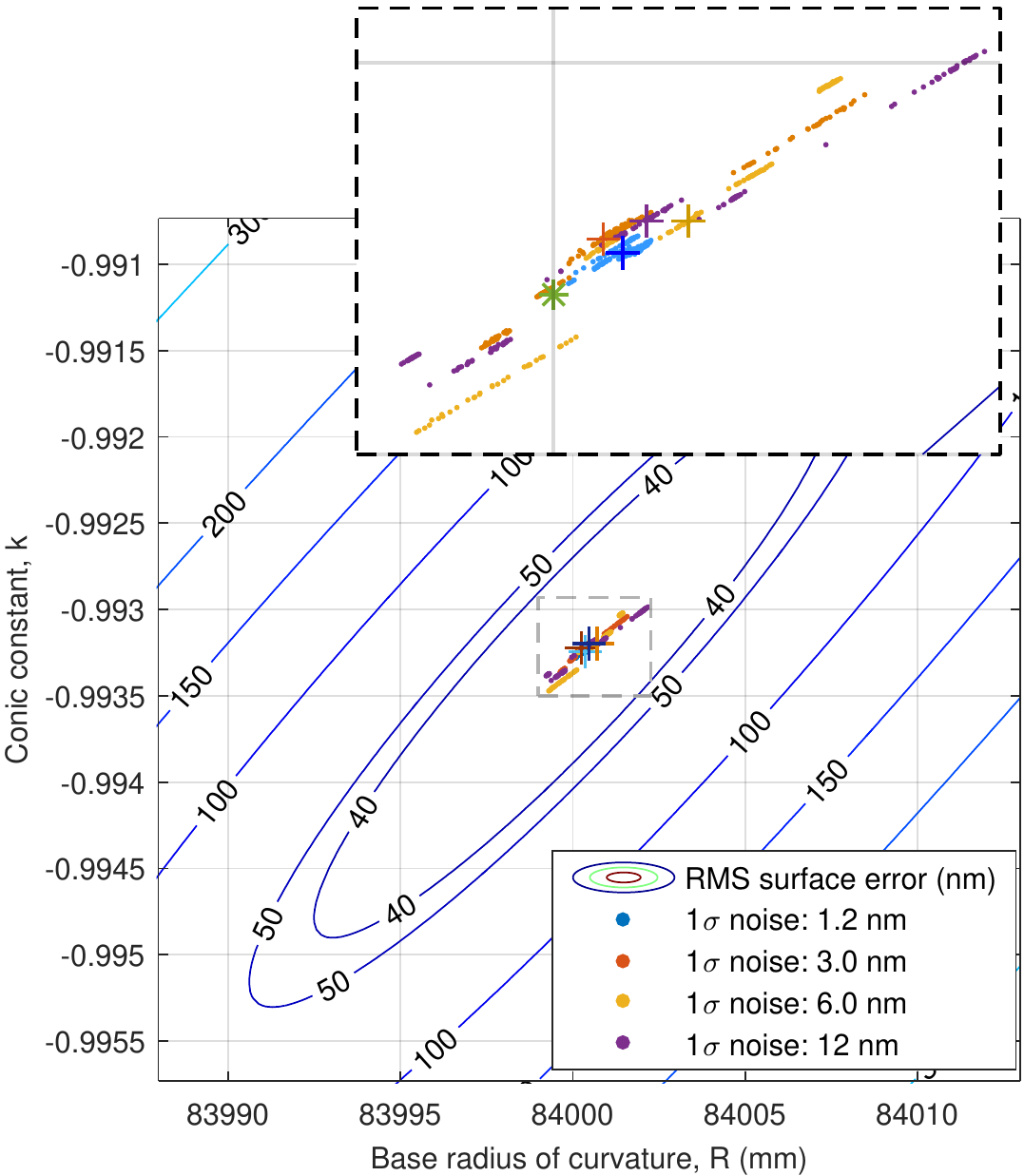}
\caption{Plot of results of 4 data groups with different random noise levels added to the data.  Each group contains 5 separate datasets with a different randomly generated noise of the level specified within that group.  Each dataset is generated using the same $R$ and $k$ (indicated by the green star on the inset top plot) and at the same outer segment position (S1).  Each dataset is analysed using the same set of 32 $(R,k)$ points as initial values to demonstrate the multi-start strategy.  The inset plot at the top is a zoom-in on the central region of the main plot; the median result of each group is indicated with a "+" marker.  The main plot is shown to demonstrate that all the results converged within the inner contour (40 nm), and within the ESO allowable error.}
\label{fig:diffnoise}
\end{figure}

\subsection{Datasets with the same fixed R and k}
\label{sec:fixRk}

\subsubsection{Datasets of varying random noise}
To investigate how the fitting routine behaves in the presence of different levels of noise, datasets were generated from a surface of a fixed base radius of curvature ($R$) and conic constant ($k$) (as per equation \ref{eq:ellipform}) with only the 1-sigma level of random noise on the $z$ height data varying between the datasets.  For these experiments: $R = 84000$mm and $k = -0.993295$ and the 1-sigma level of random noise on the $z$ data = 1.2 nm, 3.0 nm, 6.0 nm and 12 nm.  These noise levels are based around the maximum measurement error of the OpTIC NOM autocollimator which is $\pm 0.25$ arcsec and random position error of $\pm 1 \mu$m for a single point measurement.  For a 5mm step size between measurements, this angle measurement error converts to a z error of 6.0 nm (i.e. $dz = dx\tan\theta$).  In a real-life measurement 25 autocollimator measurements are acquired in rapid succession for each data point reducing this error to 1.2 nm.  The position uncertainty converts to a $z$ error of approximately 4 nm (varies depending on segment and position on segment).  Setting the 1-sigma level of z-noise in the simulations to 6.0 nm is close to an expected maximum error level and using 3.0 nm represents a more modest level of noise.  The lower and upper limits of noise (1.2 and 12 nm) have been chosen to represent an environment/set-up of excellent stability and accuracy and a more noisy/less stable environment than we assume are typical of the OpTIC NOM.

The data generated is for S1 which is positioned near the outer edge of M1 and so is the most off-axis ellipsoid.

Using the scan pattern as previously described, 5 datasets were generated using each of the 4 specified levels of random noise.  A randomised $(R, k)$ Sobol set containing 32 points was generated over an R-range of 83000 to 85000 mm and a k-range of -1.1 to -0.8.  These 32 points provide the initial $R$, $k$ values for the multi-start strategy; the same set of initial values was used to start the fitting/optimisation process for all the datasets.  Every dataset is run through the fitting process 32 times, each time using a different initial value for $R$ and $k$ as provided by the Sobol set.  The results of the final fitted $R$, $k$ for each of these datasets is shown in Fig. \ref{fig:diffnoise}, over-plotted on the contour map of $R$-$k$-RMS surface error.  The median result for each of the different noise groups is indicated.  The best-fit $R$, $k$ to the simulated NOM data is taken as this median value.  The median, rather than the mean is used as it is a better representation of the majority of the fitting results since we know that some $R$, $k$ initial values will have been poorly placed in the gradient space and that there will be a few spurious/outlier results.

\begin{figure*}[bt]
\centering
\includegraphics[width=\linewidth]{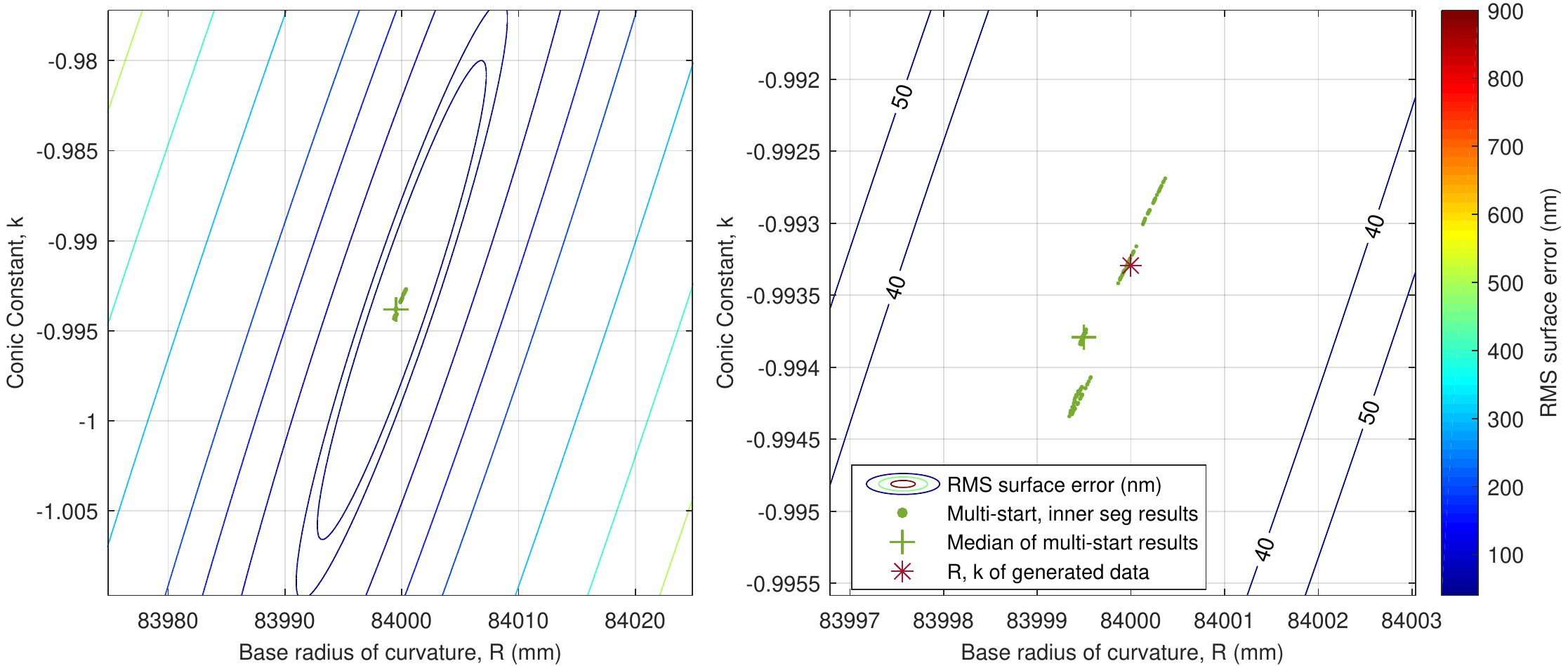}
\caption{Results of a multi-start optimisation for 5 datasets, 1-sigma noise set at 3.0 nm, same $R$, $k$ as previous but for a segment near the inner (rather than outer) edge of M1.  The $R$-$k$-cost function is different for different centre coordinates; nearer the inner edge the allowable range of $R$,$k$ values that encompass the 50 nm RMS allowable surface error is larger than those for a segment near the outer edge.  The $R$ range is similar, but the $k$ range is an order of magnitude larger at the inner edge.}
\label{fig:innerseg}
\end{figure*}

\subsubsection{Dataset at a different M1 location}
\label{sec:innerseg}
Using the same $R$ and $k$ (84000 mm and -0.993295), a z-noise level of 3.0 nm and the same set of 32 $(R, k)$ initial values, the fitting process was repeated on 5 datasets generated at a different M1 centre coordinate in order to see if there are any differences in fitting data from the inner edge of M1 i.e. closer to the edge of the central cut-out and the least off-axis ellipsoid compared to the outer edge of M1 (S1).  The centre coordinate of S1 (outer edge) in the M1 frame is [0, 18469.942] mm and of this (inner edge) segment is [0, 6393.5] mm.  As described in Section 4, a new contour map of $R$-$k$-RMS surface error was generated to compare results at this new M1 mirror position.  The results are shown in Fig. \ref{fig:innerseg}.

\begin{figure}[htbp]
\centering
\includegraphics[width=\linewidth]{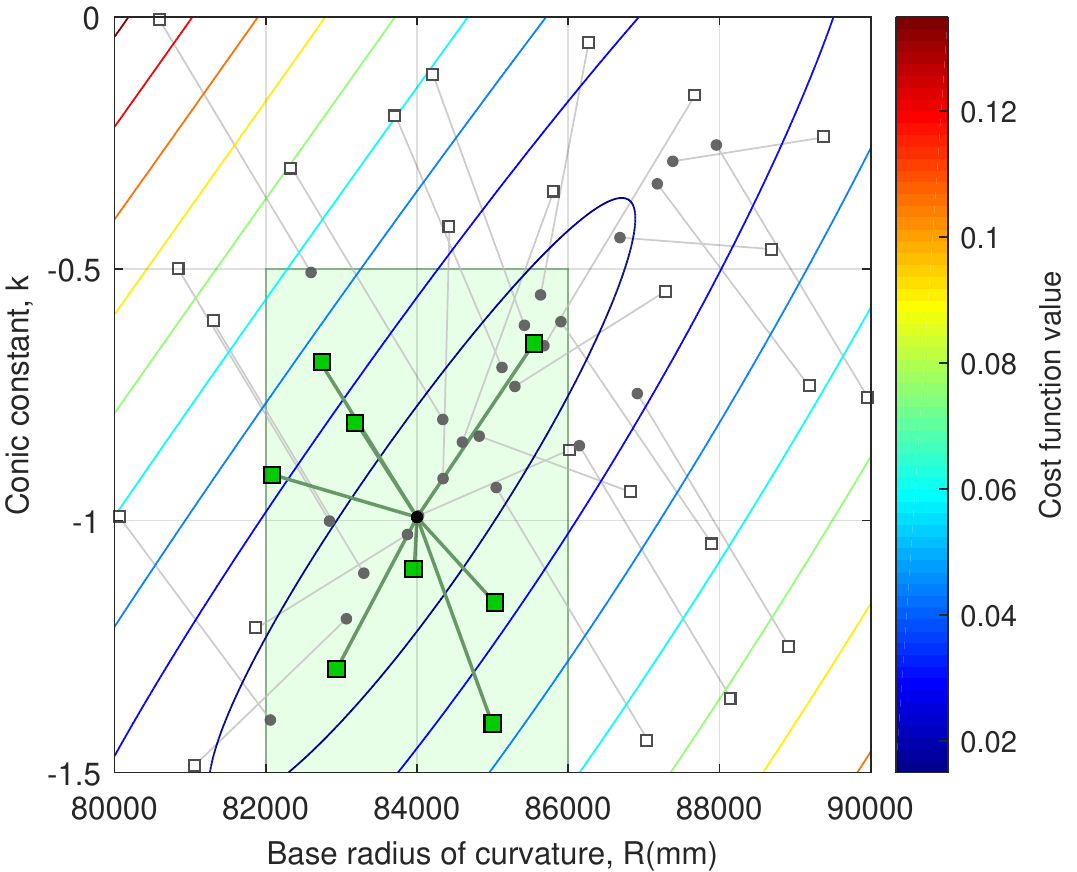}
\caption{A contour plot showing the cost function vs. fitted $R$ and $k$ values (lower cost function value indicates a better fit) given an over-wide search space for the multi-start optimisation strategy.  The square outlined points show the initial values that were passed to the optimiser for this dataset; the green shaded rectangular area indicates the boundary limit around the correct solution – initial values outside of this boundary will not be able to converge to the correct solution. The small grey points show the end-point $R$,$k$ values of the optimiser with a light-grey line linking it to its initial value.  For the 8 start values within the boundary (filled green squares) the optimiser reached the correct solution (on the contour plot there are 8 almost overlapping points at (84000, -0.993295)).}
\label{fig:extrange}
\end{figure}

\subsubsection{Dataset analysed using out-of-bounds initial values}
\label{sec:outbounds}
An extended range 32 point Sobol set of initial $(R, k)$ values was generated to demonstrate the results when the routine is provided with initial values that are outside the boundary limits around the real solution.  These boundary limits are set within the optimisation function to be: $\pm 2000$mm around $R_i$ and $\pm 0.5$ around $k_i$.  The extended Sobol set was generated using a Rrange of 80000 to 90000 and a krange of -1.5 to 0.

The NOM dataset for testing was generated with the same fixed $R$ and $k$ as previous (84000 mm, -0.993295), z-noise = 3.0 nm and at the S1 position.  The results of fitting this dataset for each of the initial values in this extended range set is shown in Fig. \ref{fig:extrange}.  It demonstrates the use of the multi-start strategy not only to avoid local minima issues, but also as a tool to “zoom-in” on the correct answer even if there is very poor knowledge of the shape of the optic from which the data was taken.  Firstly the results with the lowest value of the cost function will give a good indication of $R$, $k$ values that are at or approaching the correct results.  Also, as in the case shown in Fig. \ref{fig:extrange}, there are 8 initial values that lie within the boundary box centred on the correct $R$, $k$ value and so there is a “point” on Fig. \ref{fig:extrange} that consists of 8 almost overlapping results that started from those points and have converged on the correct solution.  Once an indication of the correct result is gleamed from an “over-wide search”, the optimisation can be repeated with a smaller range of $R$ and $k$ around the region where the correct solution is believed to lie.

\begin{figure*}[tbh]
\centering
\includegraphics[width=\linewidth]{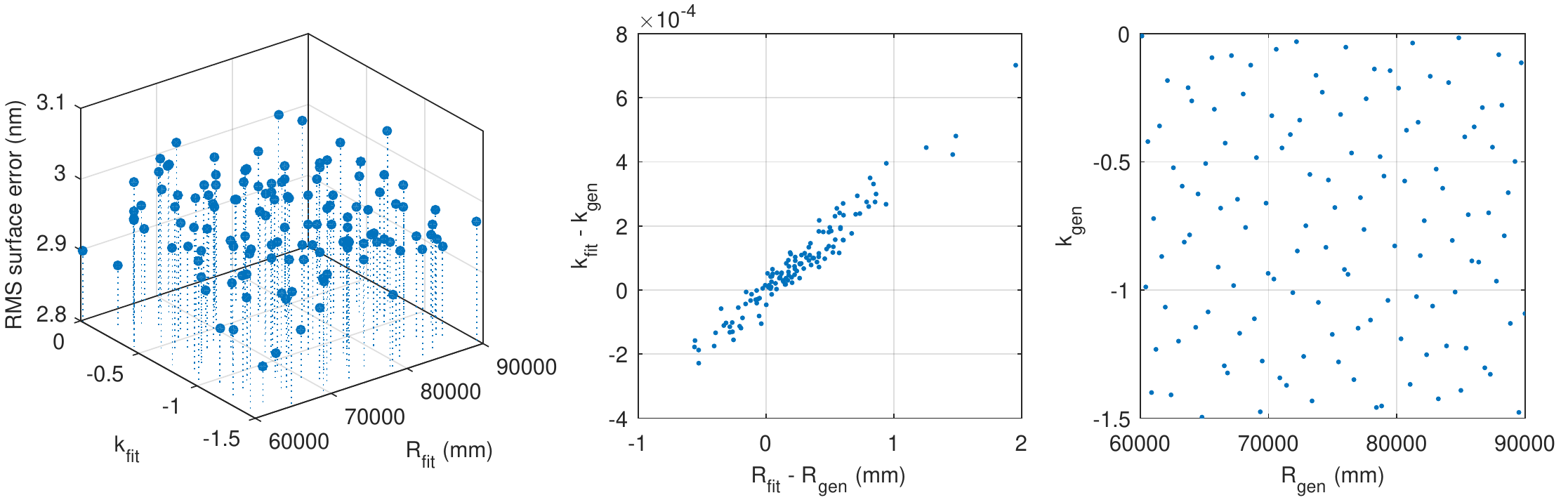}
\caption{Fitting results for 128 datasets with different $(R, k)$ surface parameters at a segment position given by cM = [0, 12000] mm. The left graph shows a plot of the fitted $R$ and $k$ values versus the residual RMS surface error. The centre plot gives another indication of how well the optimiser converged on the correct answer: plotting the difference between the fitted and actual $R$, $k$ values.  The right plot shows the $R$, $k$ pairs used to generate the data sets.}
\label{fig:varRk}
\end{figure*}

\subsection{Fixed noise (3.0 nm), varying R and k datasets}
A 128-point set of $(R, k)$ pairs was generated using a randomised Sobol set and used to generate 128 different NOM datasets.  The $R$ range for this sample was 60000 to 90000 mm and the $k$ range -1.5 to 0.  As a side-note, this range includes the current specification for the E-ELT of $R = 69000$mm and $k = -0.995882$ \cite{cayrel12}.  An arbitrary off-axis position was chosen for the simulated segment data, this was fixed for all the 128 datasets at cM = [0, 12000] mm, which is around the mid-section of a primary mirror of diameter 40000 mm.  The 1-sigma level for random noise on the data-points was also fixed at 3.0 nm.  Given the large amount of processing time that would be required to perform a many point multi-start strategy as demonstrated in Section \ref{sec:fixRk}, it was assumed for these analyses that a reasonable knowledge of the $(R, k)$ value of the mirror is known and so the initial values given to the optimiser was set to: $R_i = R_{gen} + 1000$ mm and $k_i = k_{gen} - 0.1$; where $R_{gen}$ and $k_{gen}$ are the values used to generate the simulated datasets.   The results of these optimisations are shown in Fig. \ref{fig:varRk}.  The average of the residual RMS surface error is 2.95 nm which corresponds well to the expected noise in the data (3.0 nm).  The median value of the Rdiff (fitted R value ($R_{fit}$) - R value used to generate the data ($R_{gen}$)) is 0.202 mm; the median value of the kdiff is 6.65e-5.  These are well within acceptable tolerances.

\subsection{The effect of adding form error to the datasets}
On a “real” optic, the surface form is unlikely to be perfectly described by Equation \ref{eq:ellipform} with just random noise on the data points and line-scan positions.  It is likely that the shape will be slightly modified by higher order terms such as astigmatism.  For the manufacturing of the ESO E-ELT segments there is a certain amount of allowable misfigure in terms of Zernikes \cite{eso3}, since the active mirror support structure can remove a certain amount of form error.

\begin{figure}[htbp]
\centering
\includegraphics[width=\linewidth]{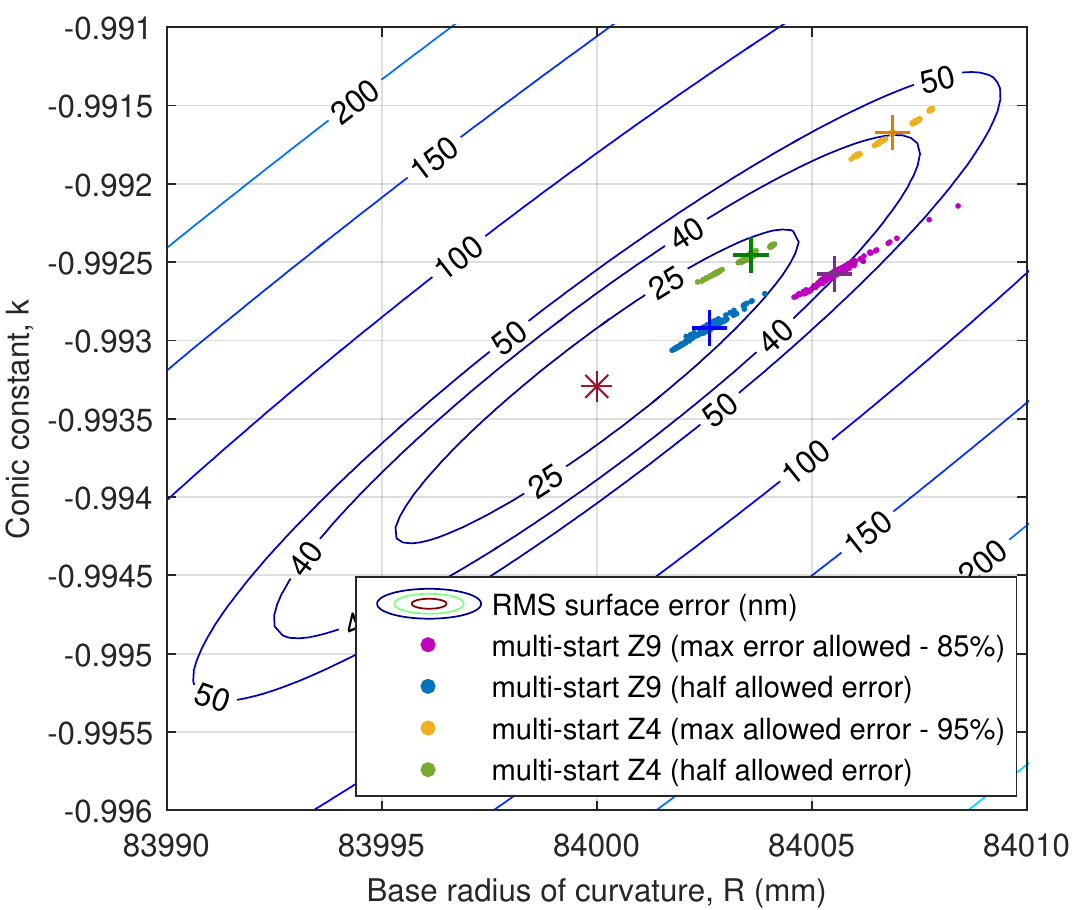}
\caption{Plot showing the results of 4 sets of multi-start optimisations.  For these data form-error was introduced to the base dataset- either astigmatism (Z4) or higher order "trefoil" (Z9), both at 2 different levels: the maximum allowed by ESO or half of this.  In all cases the medians of the fitted data sets (marked with "+" signs) fall within the allowable error boundary (50 nm contour), the asterix marks the $R$, $k$ position of the generated data.  Only 2 (out of 160) results from the maximum allowed Z9 multi-start set fall outside the boundary.}
\label{fig:formerror}
\end{figure}

For these tests we chose two of the higher order allowable errors to add to the simulated datasets to investigate how this affected the result on fitted $R$ and $k$; the misfigure in Zernike terms we added to this data are astigmatism (Z = 4) and higher order "trefoil" (Z = 9).  The misfigure terms were considered individually and were added in to the ellipsoid data either at the maximum allowed amount of error for that term or at half the allowed amount.  A 32-point multi-start strategy was used to analyse each dataset, the results are plotted in Fig. \ref{fig:formerror}.

As expected the presence of form error in the data modifies the best fit $R$ and $k$.  The presence of either astigmatism or trefoil results in a fitted $R$ and $k$ that is shifted towards a higher $R$ and $k$ than the values that were used to generate the data.  Again, as might be expected, astigmatism had a more significant effect on the fitted results.  This is partly to do with the fact that more error in astigmatism is allowed in the data but also because it can be seen how the two axes of different curvature on an astigmatic surface can add to the ellipsoid surface (also with two axes of different curvature) to produce a surface which closely resembles an ellipsoid surface with slightly modified $R$ and $k$ parameters.  Figure \ref{fig:formerror} shows that the RMS surface error due to the difference between the fitted $R$, $k$ and the actual $R$, $k$ is within allowable limits.  If the maximum allowed form errors are known to be present within the data (i.e. it is known from the interferometric data) it is advisable to subtract these form errors from the profilometer data and fit for $R$ and $k$ again to get a result with an improved accuracy.

\section{Conclusions}

This work has demonstrated that it is possible to converge on an $R$, $k$ solution of sufficient accuracy for the described purpose by using a gradient minimisation optimisation over many (100+) dimensions to place scan-lines in a common coordinate frame of reference and simultaneously fit the best aspheric surface.

The simulations performed here have assumed a fixed scan pattern of 7 line sections across x and y with data points taken every 5 mm along each of these lines (divided into 24 line scans due to the limited autocollimator measurement range).  This pattern was chosen since it represented one of the best compromises between data acquisition time and adequate sampling of the optic being measured.  The number of degrees of freedom in the optimisation is altered by the number of scan lines (6N + 2 where N is the number of scan lines).  The behaviour of the optimisation process with respect to number of scan lines has not been completely investigated.  The optimisation process is robust to a wide range of random noise on the z (height) data, a 1-sigma level of 1.2 to 12 nm has been investigated here and shown to satisfy requirements.  The process can cope with small ($\sim 10 \mu$m) random offsets of entire line scan positions since x, y, z position is part of the optimisation for each line (in order to account for line shifts when rotating the line scans into their common reference frame and the unknown integration constant).  Exact knowledge (i.e. within the noise) of the segment centre position in the data is assumed in order to have accurate mapping to the required segment centre coordinate in the M1 reference frame.  Any large or systematic errors in the coordinate system or elsewhere in the data should be corrected or removed before fitting to obtain the most accurate result; if this is not possible their influence on the fitting process will need to be investigated. 

Although the focus of these fitting tests were for the old E-ELT specification ($R = 84000$ mm, $k = -0.993295$, 42000 mm diameter primary) this fitting/optimisation process has also been shown to be effective for surfaces described by different $R$ (60000 to 90000 mm) and $k$ (0 to -1.5) and different off-axis positions (i.e. different segment centre coordinates in the M1 frame).  It also performs within specification in the presence of the allowed higher-order surface error terms (only Z4 and Z9 tested here) as defined for the E-ELT segment manufacturing, although for the best accuracy it is recommended to subtract those terms from the data before (re)fitting, especially if high levels of those errors are known to be present. 

Assuming the optic to be measured can be supported on a suitably stiff structure that allows the optic to be rotated around its x and y axes without introducing significant (i.e. below the measurement error) distortion on the surface form then the fitting method described provides a useful way to effectively extend the angular measurement range of an autocollimator in a non-contact profilometer and allows the line scans taken across the optic to be reconstructed to the best-fit aspheric (conic) surface. 

This fitting process has since been adapted for use within a modified measurement procedure using the OpTIC NOM \cite{Pearson15}, with a reduced set of optimisation parameters per line-scan.

\bigskip
\noindent \textbf{Funding Information.}  Welsh Assembly Government via the A4B (Academics for Business) Programme; EPSRC Translation Grant into Ultra Precision Surfaces (ref: EP/F031416/1).



\begin{thebibliography}{1}
\newcommand{\enquote}[1]{``#1''}

\bibitem{Atkins11}
C.~{Atkins}, J.~{Mitchell}, and P.~{Rees}, \enquote{{Non-contact profilometry
  of E-ELT segments at OpTIC Glyndwr},}  (SPIE, 2011), vol. 8126, p. 81260M.

\bibitem{Pearson15}
J.~L. Pearson, G.~W. Roberts, P.~C.~T. Rees, and S.~J. Thompson, \enquote{{Use
  of a NOM profilometer to measure large aspheric surfaces},}  (SPIE, 2015),
  vol. 9628, pp. 96280W--96280W--12.

\bibitem{qian95}
S.~Qian, W.~Jark, and P.~Z. Takacs, \enquote{{The penta-prism LTP: A
  long-trace-profiler with stationary optical head and moving penta-prism},}
  Rev. Sci. Instrum. \textbf{66}, 2562 (1995).

\bibitem{alcock10}
S.~Alcock, K.~Sawhney, S.~Scott, U.~Pedersen, R.~Walton, F.~Siewert,
  T.~Zeschke, F.~Senf, T.~Noll, and H.~Lammert, \enquote{{The Diamond NOM: A
  non-contact profiler capable of characterizing optical figure error with
  sub-nanometre repeatability},} Nuclear Instruments and Methods in Physics
  Research A \textbf{616}, 224 (2010).

\bibitem{matlab}
Mathworks, \enquote{\uppercase{M}\textsc{ATLAB} release 2015a,} Tech. rep., The
  MathWorks, Inc., Natick, Massachusetts, United States (2015).

\bibitem{eso3}
ESO, \enquote{\uppercase{E-ELT} programme, specifications for the call for
  tenders "supply of 7 prototype segments of the 42m diameter \uppercase{E-ELT}
  primary mirror,} \uppercase{E-SPE-ESO-300-0150} issue 3 (2009).

\bibitem{recipes}
W.~H. Press, S.~A. Teukolsky, W.~T. Vetterling, and B.~P. Flannery,
  \emph{Numerical Recipes, The Art of Scientific Computing 3rd Edition}
  (Cambridge University Press, 2007).

\bibitem{sobol76}
I.~Sobol and Y.~Levitan, \enquote{The production of points uniformly
  distributed in a multidimensional cube,} Tech. rep. 40, Institute of Applied
  Mathematics, USSR Academy of Sciences (1976).

\bibitem{cayrel12}
M.~Cayrel, \enquote{{E-ELT optomechanics: overview},}  (SPIE, 2012), vol. 8444.

\end{thebibliography}
\end{document}